\pdfoutput=1
\documentclass[sigconf]{acmart}

\setcopyright{none}
\renewcommand\footnotetextcopyrightpermission[1]{}

\AtBeginDocument{%
  }

\usepackage{algorithm,algpseudocode}
\usepackage{subcaption}
\usepackage{enumitem}
\usepackage{threeparttable}
\usepackage{diagbox}
\usepackage{multirow}
\usepackage{hyperref}
\usepackage{booktabs}
\usepackage{makecell}

\definecolor{colorreg}{RGB}{255, 0, 102}
\definecolor{colormne}{RGB}{0, 0, 255}
\definecolor{colorimm}{RGB}{84, 130, 53}
\definecolor{colorbyt}{RGB}{191, 144, 0}
\definecolor{colorcom}{RGB}{112, 48, 160}
\definecolor{colorrebuttal}{RGB}{255, 0, 0}

\newcommand{\clab}[1]{\textbf{\texttt{{#1}}}}
\newcommand{\creg}[1]{\textbf{\texttt{{#1}}}}
\newcommand{\cmne}[1]{\textbf{\texttt{{\color{colormne}#1}}}}
\newcommand{\cimm}[1]{\textbf{\texttt{{#1}}}}

\begin{document}

\title{(Ab)using Processor Exceptions for Binary Instrumentation on Bare-metal Embedded Firmware}

\author{Shipei Qu, Xiaolin Zhang, Chi Zhang, Dawu Gu}
\email{{shipeiqu, xiaolinzhang, zcsjtu, dwgu}@sjtu.edu.cn}
\affiliation{%
  \institution{Shanghai Jiao Tong University}
  \city{Minhang Qu}
  \state{Shanghai Shi}
  \country{China}
}

\begin{abstract}
  Analyzing the security of closed-source drivers and libraries in embedded systems holds significant importance, given their fundamental role in the supply chain. Unlike x86, embedded platforms lack comprehensive binary manipulating tools, making it difficult for researchers and developers to effectively detect and patch security issues in such closed-source components. Existing works either depend on full-fledged operating system features or suffer from tedious corner cases, restricting their application to bare-metal firmware prevalent in embedded environments. 
  
  In this paper, we present PIFER (Practical Instrumenting Framework for Embedded fiRmware) that enables general and fine-grained static binary instrumentation for embedded bare-metal firmware. By abusing the built-in hardware exception-handling mechanism of the embedded processors, PIFER can perform instrumentation on arbitrary target addresses. Additionally, We propose an instruction translation-based scheme to guarantee the correct execution of the original firmware after patching. We evaluate PIFER against real-world, complex firmware, including Zephyr RTOS, CoreMark benchmark, and a close-sourced commercial product. The results indicate that PIFER correctly instrumented 98.9\% of the instructions. Further, a comprehensive performance evaluation was conducted, demonstrating the practicality and efficiency of our work.

\end{abstract}

\keywords{Embedded System Security, Binary Rewriting, }

\maketitle

\pagestyle{plain}

\section{Introduction}

In recent years, the security of embedded systems has become increasingly important with their broad applications. Designed for low-cost environments, the majority of embedded devices/chips are \textit{bare-metal}, i.e., running only a single statically linked firmware. However, the lack of advanced operating systems and architectural security features, such as a Memory Mapping Unit (MMU), render bare-metal embedded devices particularly vulnerable to programming errors and malicious attacks \cite{alrawi2019sok}. Moreover, bare-metal chips/devices are being actively exploited in sophisticated attacks against high-value targets, such as the compromising of AMD-SP \cite{buhren2021one}, Google Pixel's security co-processor \cite{Titanm:online}, and Tesla's key-fob \cite{wouters2021my}. Therefore, it is crucial to efficiently detect and defend against vulnerabilities of such devices.

As many widely used libraries or drivers are still closed-source in the IoT-embedded industry, binary instrumentation is one of the fundamental techniques in addressing the above tasks. It involves injecting additional code into the binary program, allowing developers and security researchers to observe or modify its runtime behavior. 
From a defense perspective, binary instrumentation enables vendors to perform feedback-based fuzzing and custom security patches for low-level closed-source components that are frequently targeted in supply chain attacks. Take Ripple20 \cite{Ripple20:online} for an example, a vulnerable proprietary TCP/IP stack risks hundreds of millions of devices, and 120 days is insufficient for its developers to deliver the correct updates to all affected vendors. 
For offensive security research, the binary-only firmware also poses a hindrance to reverse engineering and exploitation. For example, hardware-based attacks, which have gained popularity in recent years, typically enable the bypass of secure boot or debugging protections \cite{GitHubst27:online}. However, researchers find it tedious to manipulate and analyze the raw binary firmware after a successful attack. In general, efficient modification at the binary level is crucial for both offensive and defensive security purposes. 

\textbf{Related work} While many static binary rewriting techniques work well on x86,
limited studies has been conducted for embedded architectures\cite{kim2017revarm}\cite{salehi2020musbs}\cite{diarmore}. Existing binary instrumentation techniques can be classified into \textit{static} and \textit{dynamic}, depending on whether the modifications to the program are made before or during execution. For static approaches, the most common trampoline way replaces an original instruction with a \textit{jump} targeting the newly appended code. However, in embedded architectures, the jump or addressing range of a single instruction is quite limited due to the fixed length (i.e., 2 bytes for ARM-Thumb/RISCV-Compressed), failing many rewriters commonly worked on x86 \cite{hunt1999detours}\cite{deng2013bistro}. Another approach for static rewriting is to lift the entire binary to mutable Intermediate Representation (IR) or even re-assemble assembly \cite{dinesh2020retrowrite}, but a recent study reveals that related works are error-prone in real-world applications, as generically distinguishing between pointers and scalars is still an undecidable problem in binary analysis \cite{kimreassembly}. 

For dynamic approaches, using hardware debugging protocols (JTAG or SWD \cite{du2022afliot}) to modify firmware logic will introduce significant communication overhead and rarely be deployed in released production. Rehosting under emulated environments is also a potential dynamic approach \cite{eisele2022embedded}, but limited scalability and poor performance may arise. 
Another dynamic solution is the trap-based approach, which works by modifying the target instructions to trap raisers (e.g., the \clab{INT3} in x86) and performs the instrumentation with pre-registered handlers. 
Traditional trap-based approaches like GDB require the restoration of modified instructions to ensure correctness. However, the ROM/FLASH memory where firmware typically resides does not allow efficient on-the-fly modifications at the byte level. For specific tasks like control flow integrity \cite{nyman2017cfi}, a workaround is to manually emulate a small set of instructions (call, return) in other locations. 
However, for general instrumentation, 
it is challenging to correctly handle all possible 16/32 bits instructions. 
Furthermore, bare-metal firmware lacks a comprehensive operating system like Linux that enables the registration of custom handlers, and integrating an efficient self-contained hook system into the compact binary firmware is also a challenging task.

\textbf{Contributions}  
In this paper, we propose PIFER, a fine-grained and practical binary instrumenting framework designed for bare-metal firmware. 
In general, PIFER combines the advantages of the dynamic trap-based rewriting and the static patching. 
Utilizing the exception-handling mechanism of embedded processors, 
PIFER achieves execution flow hijacking within a compact 2-byte (1-\textit{WORD}) sequence, which is fitted into fixed-length instructions and thus applicable at arbitrary addresses. 
To ensure correct re-execution of the modified original instructions, we design a novel
single instruction translator capable of converting arbitrary regular instructions into relocated equivalents. 
Overcoming the limitations of traditional binary instrumenting within embedded firmware, 
PIFER offers a concrete solution to facilitate late-stage code modifications on bare-metal firmware for both security and developmental purposes. 
Furthermore, a comprehensive evaluation demonstrates that PIFER is able to handle real-world complex firmware with acceptable runtime and memory overhead.

To summarize, our core contributions are:

\begin{itemize}
    \item We present the first general and fine-grained static binary instrumenting framework for bare-metal embedded firmware. 
    \item We propose an instruction translation scheme that transforms trap-based rewriting into a static method, enabling its application on embedded devices. Combining it with the abusable exceptions we identified, we build a practical hooking system focusing on bare-metal embedded firmware. 
    \item We develop and open source\footnote{\url{https://github.com/PIFER-release/PIFER}} an out-of-box prototype of PIFER for ARM supporting all Cortex-M microprocessors, which are the most widely used architectures in IoT industry.  
    \item We conduct a comprehensive evaluation on both benchmarks and real-world firmware under diverse devices, as well as a discussion of the applicability and limitations of PIFER.
\end{itemize}

\section{Background and motivation}

In this Section, we discuss binary rewriting for embedded firmware and the limitations of existing work.

\subsection{Binary Rewriting for Embedded Firmware}

The firmware of an embedded device refers to the software stored in non-volatile memory and is responsible for controlling the application logic of specific hardware, such as microcontrollers. Crafted by different vendors for dedicated tasks, the majority of such firmware is proprietary and not open-source. This situation poses a challenge to both security researchers and downstream developers in the embedded ecosystem, as several important applications require modifications of the target, such as bug fixes, fault observability, performance profiling, and custom feature additions \cite{salehi2020musbs}. Direct binary rewriting could accomplish the above tasks without source code but requires extremely delicate construction to avoid breaking the original code and causing errors. Unfortunately, for bare-metal embedded firmware, existing binary rewriters all have limitations in both methodology and implementation.

\begin{figure}
  \centering
  \begin{subfigure}{0.80\linewidth}
      \centering
      \includegraphics[width=1.0\linewidth]{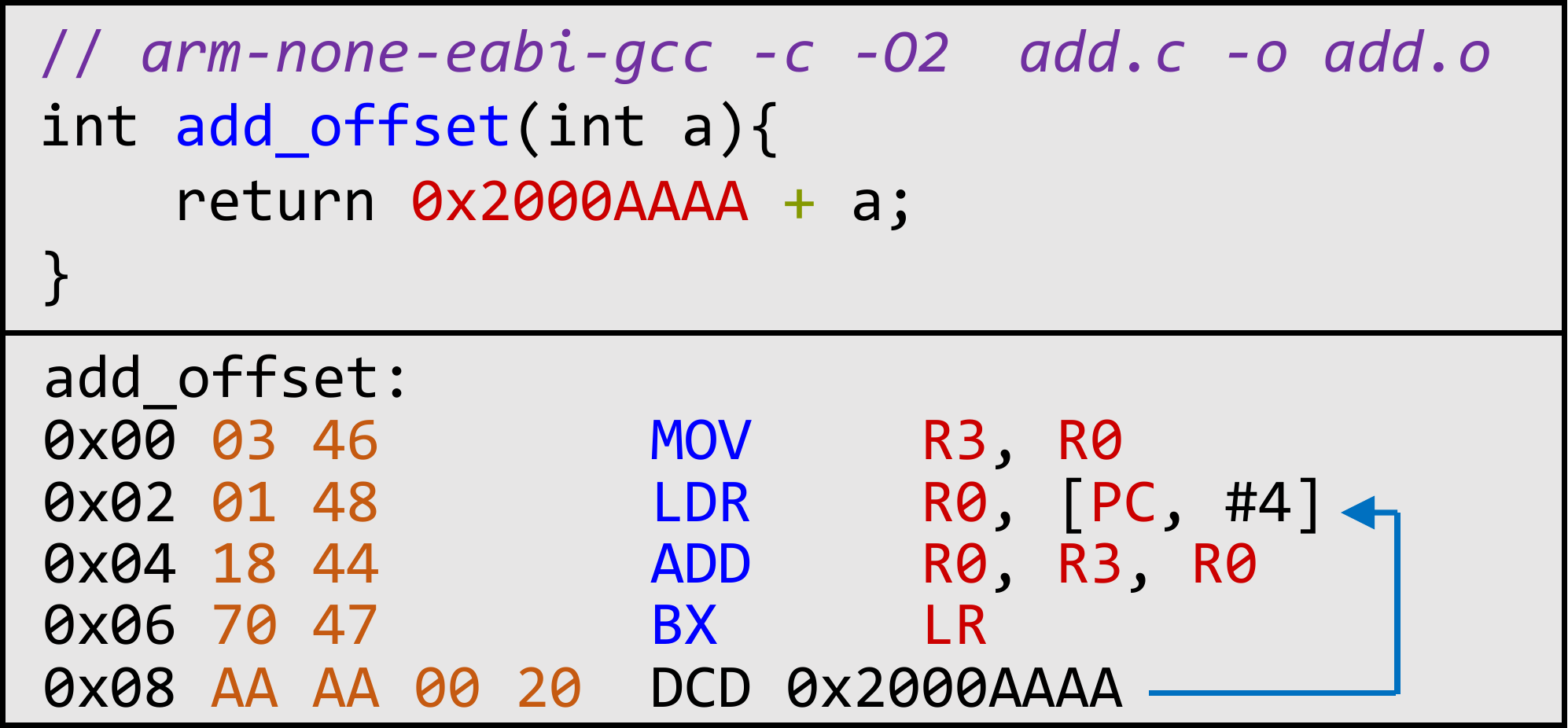}
      \caption{Mixed code/data in ARM.}
  \end{subfigure}
  \begin{subfigure}{0.80\linewidth}
      \centering
      \includegraphics[width=1.0\linewidth]{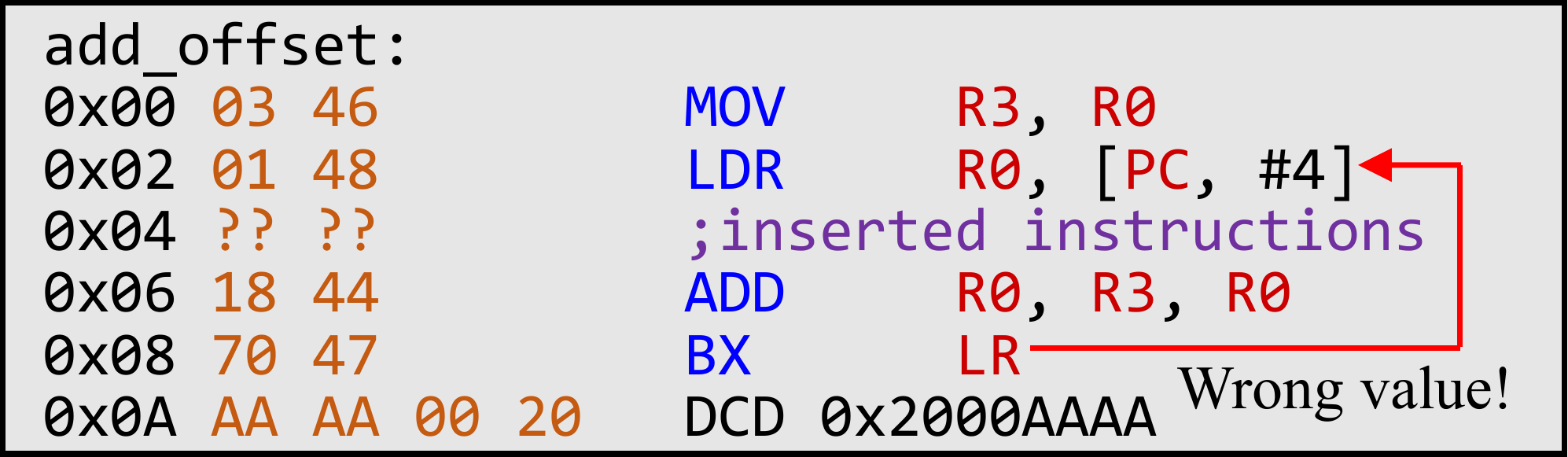}
      \caption{After instrumenting.}
  \end{subfigure}
  \caption{Example of table-based approach failure on ARM. A PC-relative addressing instruction (at \textbf{0x02}) is used to fetch data from the literal pools (at \textbf{0x10}). After inserting the new code into this basic block, the offset between the literal pools and the original addressing instruction is changed and an unexpected value (\cimm{0x47704418}) is loaded into \creg{R0}.}
  \label{fig:sbs}
\end{figure}

\subsection{Limitations of Existing Approaches}

\label{subsec:limitations}

Existing static binary rewriting techniques can be classified into three categories\cite{kimreassembly}: (i) detour-based; (ii) table-based; (iii) symbolization/reassembly.

\noindent \textbf{Detour-based Instrumentation} Detour-based instrumentation techniques, such as Detours\cite{hunt1999detours}, Bistro\cite{deng2013bistro}, and E9Patch\cite{duck2020binary}, overwrite the original instruction at the target location to a \textit{jump} which redirects the control flow to a \textit{trampoline} consisting of newly inserted instructions. 
As identified in \cite{kim2017revarm}, this approach struggles to work on embedded architectures due to the short jump range of fixed-length instructions. 
Take Cortex-M for an example, when the trampoline is more than 2048 bytes away from the target, a long jump (\cmne{B.W}, consuming 4 bytes) has to be used, which will not fit into the 16-bit Thumb instructions.

\noindent \textbf{Table-based Instrumentation} Table-based approaches such as Multiverse\cite{bauman2018superset}, AflIot\cite{du2022afliot}, $\mu$SBS\cite{salehi2020musbs}, and ARMore \cite{diarmore}, restore address offsets corrupted by instrumented binary code by redirecting control flow transfer instructions to a pre-computed mapping table. However, this approach assumes that data references remain static. As illustrated in Figure \ref{fig:sbs}, the table-based approach fails to handle the literal-pools feature in embedded architectures like ARM, where the data and code could be mixed in \texttt{.text} segment for memory optimization \cite{ARMLiteral-pools:online}. ARMore is the only work that considers mixed data/code, but their segfault-catching workaround cannot be applied on bare-metal firmware.

\noindent \textbf{Reassembly} Reassembly-based technologies such as RetroWrite\cite{dinesh2020retrowrite}, work by translating the binary into a relocatable IR or even the equivalent assembly produced by compiler\cite{kimreassembly}. Specifically, the aim is to generate the equivalent assembly code with symbolic labels as the compiler produces. 
However, recovering symbolic labels from the binary is difficult in theory. As pointed out by recent work\cite{kimreassembly}, all existing reassembly-based frameworks fail to achieve the soundness they claimed.

\section{Design}

Our aim is to perform \textit{practical} static binary instrumentation for bare metal firmware of embedded devices, with the following requirements:

\begin{enumerate}[leftmargin=*]
    \item \textbf{Sound:} The firmware must be working correctly after patching.
    \item \textbf{Fine-grained:} Instrumentation at arbitrary binary positions.
    \item \textbf{Scalable:} Methodologically covers as many embedded devices as possible. 
    \item \textbf{Efficient:} Fast instrumentation of the original firmware without excessive human effort and knowledge. 
    \item \textbf{Real-world Usable:} Works out of the box in the raw binary-only context (i.e., without any compiler features or debugging information), which is typical in real-world analysis.
\end{enumerate}

In the rest of this Section, we will first introduce the overview of PIFER's workflow together with the structure of the hooking system, and then dive into the details that support them. 

\subsection{Overview}

\begin{figure}
  \centering
  \includegraphics[width=0.65\linewidth]{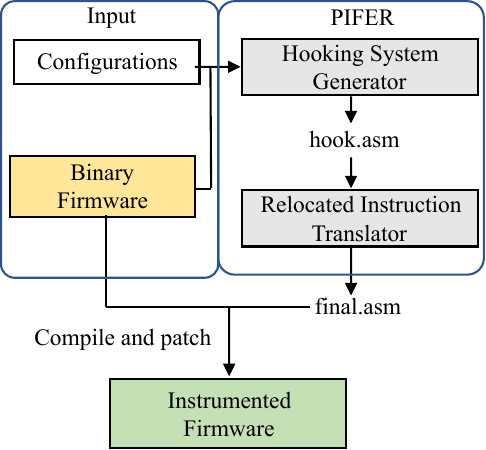}
  \caption{Overview of PIFER's workflow.}
  \label{fig:workflow}
\end{figure}

As illustrated in Figure \ref{fig:workflow}, the workflow of PIFER can be
broken down into three steps:

\begin{itemize}[leftmargin=*]
  \item \textbf{Step 1: Generating the hooking system.} The input of PIFER consists of the target binary firmware and a configuration file. The configuration contains information about the base address and architecture of the firmware, the offset of the exception vector table (EVT)\footnote[1]{We assume that these attributes are already known to the user, which is the most basic step in analyzing a binary firmware and is facilitated by tools like Binwalk\cite{GitHubRe13:online}.}, and the instrument targets with new code to be inserted. 
  PIFER will then parse the raw firmware and generate a self-contained hooking system in assembly (Section \ref{sec:hook-sys}). 
  \item \textbf{Step 2: Translating the modified instructions.} Next, PIFER will translate the instructions at instrumenting locations to their relocated equivalents and incorporate them into the hooking system thus ensuring correctness (Section \ref{sec:proxy-rewriter}). 
  \item \textbf{Step 3: Binary patching.} In this step, the assembly generated in Step 2 is compiled and appended to the original firmware. Next, PIFER modifies the addresses of the selected exception handler in the vector table according to the configuration input (Section \ref{sec:binary-gen}).
\end{itemize}

\subsection{The Hooking System}
\label{sec:hook-sys}

\begin{figure}
  \centering
  \includegraphics[width=0.9\linewidth]{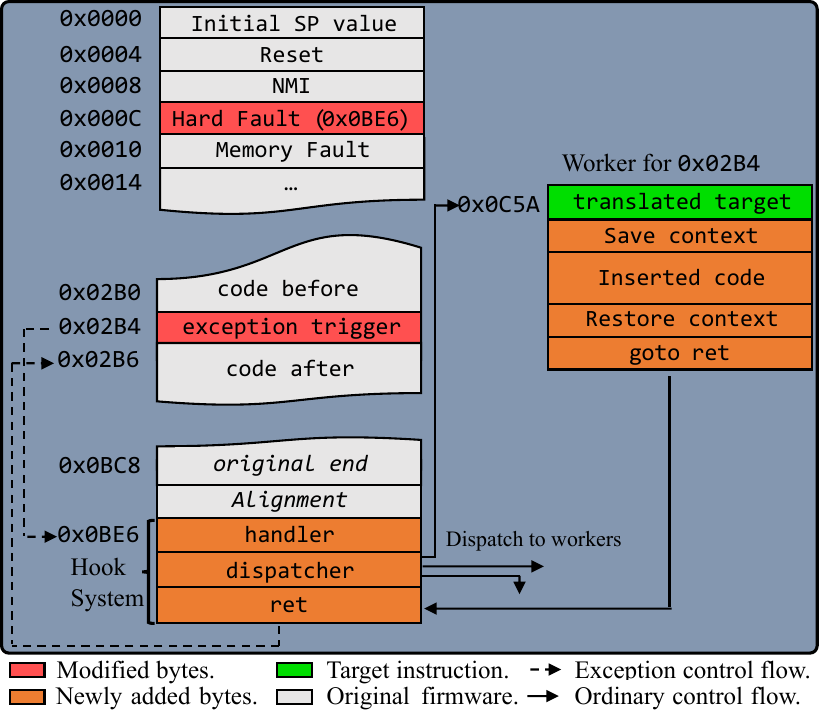}
  \caption{The designed hook system for ARM architecture.}
  \label{fig:hook-sys}
\end{figure}

Driven by the processor exception handling mechanism, PIFER constructs a self-contained hooking system as illustrated by the example in Figure \ref{fig:hook-sys}. The instruction at the target address is replaced with an exception-raising one and the corresponding handler is modified to the handler of the hook system (detailed in Section \ref{sec:binary-gen}). As the program executes to the target address instruction, it will be redirected to the following procedure:

\begin{itemize}[leftmargin=*]
  \item \textbf{handler:} This step will save the information in the exception context, including the exception occurred address and the registers state. It will determine whether the exception was intentionally raised by the PIFER, and thus choose whether to continue to the \clab{dispatcher} or fall back to the original exception handler. 
  \item \textbf{dispatcher:} In this step, according to the address where the exception occurred, a jump will be made to the corresponding instrument code, which is referred to as \clab{worker} in our design. 
  \item \textbf{worker:} The worker will first execute the translated instruction, thus ensuring that the overwritten instruction (the original one at \clab{0x02B4} in Figure 3)  still executes correctly. It then saves the program context and executes the newly inserted instructions. Afterward, it restores the context and returns to the \clab{ret} module. 
  \item \textbf{ret:} Finally, we update the new context to the stored pre-exception one and return to the original next instruction. 
\end{itemize}

Thus, we build a self-contained hooking system that operates simply by utilizing the exception-handling mechanisms common in most embedded architectures. 
However, as we discussed earlier, there are two critical challenges to make it truly operational: (\textbf{RC1}) correctly re-execution of the overwritten instruction, and (\textbf{RC2}) arbitrarily triggering desired exception within 2 bytes. 

\subsection{Translating the Overwritten Instructions}
\label{sec:proxy-rewriter}

Considering the combination of different operands and operations, the number of valid 16/32 bits instructions is quite large. Therefore, it is a tedious challenge to correctly handle all the possible overwritten instructions for general fine-grained rewriting. To address it, we first group all possible instructions into the following two classes: 

\begin{itemize}[leftmargin=*]
    \item \textbf{C1: Instructions not involving \creg{PC} register.} For example, arithmetic operations and data transfer instructions. 
    \item \textbf{C2: Instructions involving \creg{PC} register.} Examples include control flow-related instructions (e.g., jumps, call to/return from functions), and PC-relative addressing. 
\end{itemize}

For class C1, relocation of these instructions does not impact the runtime effect, thus simple replay suffices. 
For instructions in C2, we propose a \textit{register proxy} method to build transformed instructions producing the same effect as the original ones. 
The pseudo-code in Algorithm \ref{algo:constructc2handler} describes how PIFER constructs the assembly code to handle C2 instructions from a high-level perspective, taking the target instruction, its location \creg{context}.\textit{PC}, and the address of the original next instruction \creg{context}.\textit{RA} (i.e., where the handler should return after processing the exception). 
Specifically, we first convert all C2 instructions into their equivalents but \textit{explicitly} using the PC register (line 4). For example, the PC-relative address generating instruction in ARM \cmne{ADR} \creg{Rd} \clab{label} can be transformed to \cmne{MOV} \creg{Rd}, \creg{PC;} \cmne{ADD} \creg{Rd}, \creg{Rd}, \clab{label}. 
In our implementation for ARM architecture, the rest C2 instructions that non-explicitly use the PC are transformed with the following rules:

\begin{itemize}[leftmargin=*]
    \item \textbf{B, BL, BX, and BLX}: The branch instructions can be universally represented as \cmne{B\{L\}\{X\}\{cond\}} \clab{label}/\creg{Rn}. Noting that it is actually a conditional assignment to the PC instruction, we can use the \cmne{B\{cond\}}+\cmne{LDR}+\cmne{MOV} instruction to translate it. For the branch with link instruction \cmne{BL}, an additional step in the handler is required to set the \creg{LR} register in the saved context to the address of the next instruction. 
    \item \textbf{CBZ, CBNZ}: The \cmne{CBZ/CBNZ} instruction does not change condition flags but is otherwise equivalent to a combination of \cmne{CMP}+\cmne{BEQ}. We apply this solution directly by saving all flags (\creg{xPSR}) on the stack before execution and restoring them afterward.
    \item \textbf{TBB, TBH}: The \cmne{TBB} [\creg{Rn}, \creg{Rm}] instruction can be broken into two steps: 1) Get a byte at the address of \creg{Rn}+\creg{Rm}. 2) Add the byte to the \creg{PC}. Correspondingly, our translation can be split into two parts: first use the proxy register \creg{Rx} to get the value of that byte(\cmne{LDRB} \creg{Rx}, [\creg{Rn}, \creg{Rm}]), then add its value to the \creg{PC} stored in the context by a dedicated \clab{Worker} under handler mode. The same strategy also applies to \cmne{TBH}, which is just a half-word version of \cmne{TBB}.
    \item \textbf{IT}: The If-Then instruction can control up to 4 subsequent instructions. However, it does not directly alter the control flow but rather informs the processor whether each controlled instruction should take effect based on the flag bits. Since the \creg{xPSR} storing the flags is automatically saved in the exception stack, we can simply treat the \cmne{IT} as a PC-independent instruction.
\end{itemize}

Then, we search for a register \clab{rx} that is not used by the overwritten instruction (lines 5-10). The searching implementation could be improved in the future as it may fail when a single instruction occupies \textit{all} available registers. 
However, we did not find any such instructions in real-world firmware in our experiments.
Next, we replace the literal \texttt{"PC"} in the translated assembly with the \clab{rx} (line 11,14) and add the exchange code between \creg{rx} and \creg{context}.\textit{PC} in the saved context before and after it (lines 12-13, 15-16). The sub-procedures \textsc{SaveRx}, \textsc{RestoreRx}, \textsc{ContextReadPC}, \textsc{ContextWriteRA} will automatically generate the assembly code for swapping in and out the content of \creg{rx} and \creg{context}.\textit{PC}/\creg{context}.\textit{RA}. 
In other words, we use \creg{rx} as a \textit{proxy} for the original \creg{PC}, thus avoiding the incorrect execution after relocation thereby addressing \textbf{RC1}.

\begin{figure}[t]
    \centering
    \includegraphics[width=0.8\linewidth]{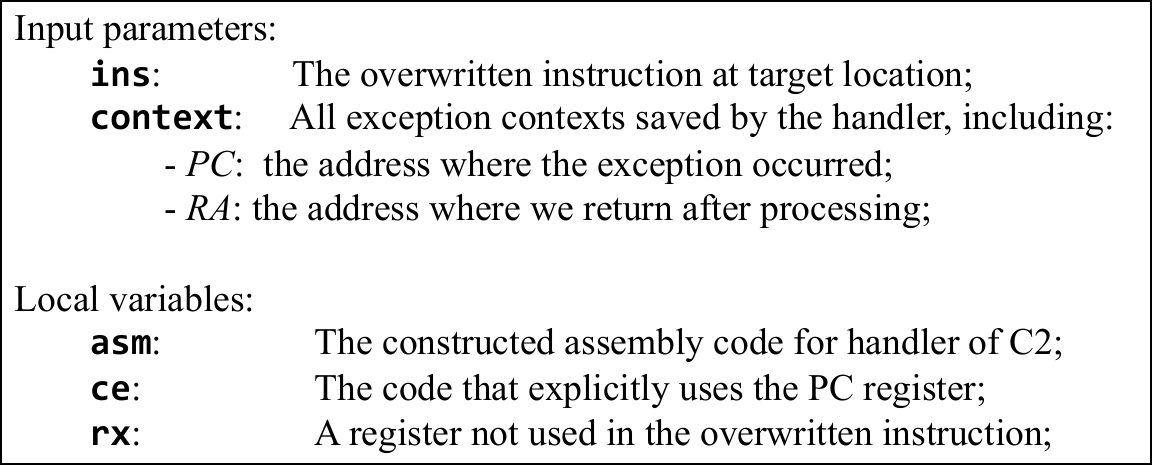}
    \caption{Definitions of variables used in Algorithms 1.}
    \label{fig:def-vars1}
\end{figure}

\begin{algorithm}[t]
    \caption{Generating equivalent assembly for C3 instructions}
    \scriptsize
    \label{algo:constructc2handler}
    \begin{algorithmic}[1]
        \Function{ConstructC3Handler}{\creg{ins}, \creg{context}}
        \State \creg{asm} := empty string
        \State \creg{rx} := Null
        \State \creg{ce} := \Call{Translate}{\creg{ins}} \algorithmiccomment{Translate to assembly that explicitly using \creg{PC} register.}
        \For{\creg{r} in \texttt{\{All general registers\}}} \algorithmiccomment{Find a proxy register not used by \creg{ce}.}
            \If{\creg{r} not in \creg{ce}}
            \State \creg{rx} := \creg{r}
            \State \textbf{break}
            \EndIf
        \EndFor
        \State \creg{ce} = \Call{Replace}{\creg{ce}, \texttt{"PC"}, \creg{rx}} \algorithmiccomment{Replace all \texttt{"PC"}s in \creg{ce} with \creg{rx}.}
        \State \creg{asm}.\textit{append}(\Call{SaveRx}{\creg{rx}}) \algorithmiccomment{Backup the content in \creg{rx}.}
        \State \creg{asm}.\textit{append}(\Call{ContextReadPC}{\creg{context}.\textit{PC}, \creg{rx}}) \algorithmiccomment{Overwrite the content in \creg{rx}.}
        \State \creg{asm}.\textit{append}(\creg{ce}) \algorithmiccomment{Translated target instruction.}
        \State \creg{asm}.\textit{append}(\Call{ContextWriteRA}{\creg{rx}, \creg{context}.\textit{RA}}) \algorithmiccomment{Set the new target address.}
        \State \creg{asm}.\textit{append}(\Call{RestoreRx}{\creg{rx}} ) \algorithmiccomment{Restore the original content in \creg{rx}.}

        \State \textbf{return} \creg{asm}  \algorithmiccomment{return the constructed assembly code.}
        \EndFunction
    \end{algorithmic}
\end{algorithm}

\subsection{Exception Raising and Binary Patching}
\label{sec:binary-gen}

The hook system in Section \ref{sec:hook-sys} requires patching the target instruction with a designated exception-causing one and subsequently adjusting the corresponding EVT handler function. We conduct a thorough research and finally choose the \textit{illegal instruction} exception for the following reasons:

\begin{itemize}[leftmargin=*]
  \item \textbf{Size:} A modification within only 2 bytes is enough to make a normal instruction illegal.
  \item \textbf{Universal:} Architectures such as ARM and RISCV have "standard" undefined instructions that can reliably trigger exceptions. For example, the opcode pattern \clab{0x00 0xDE} will raise an exception across ArmV6-M, Armv7-M, and
  Armv8-M devices.
  \item \textbf{Priority:} The undefined/illegal instruction exception typically has a high priority and could be further elevated to \clab{Hard Fault}, which is desirable since no regular exceptions or interrupts may preempt the hooking system. 
\end{itemize}

After modifying the target instruction and corresponding handlers in the EVT, PIFER compiles the assembly produced by the previous steps to binary form and appends it to the original firmware. The layout of the instrumented firmware is also illustrated in Figure \ref{fig:hook-sys} (best viewed in color).

\section{Evaluation}
\label{sec:eval}

Our evaluation was conducted to answer the following research questions that support our early claims of PIFER:

\begin{itemize}[leftmargin=*]
    \item \textbf{Correctness}: Is the behavior of the firmware consistent before and after the instrumentation?
    \item \textbf{Performace}: Does its runtime and memory overhead suffice for security applications such as vulnerability patching? 
    \item \textbf{Scalability}: Does it scale to real-world, complex firmware?
\end{itemize}

To answer the above questions, we implement a prototype of PIFER for the ARM architecture using about 2,000 lines of Python code and carry out diverse measurements and case studies.

\subsection{RTOS instrumentation on multiple platforms (Correctness, Scalability)}

\noindent \textbf{Experiment settings.} To investigate the scalability of PIFER, we perform instrumenting experiments against a non-trivial, representative embedded firmware: the Zephyr RTOS \cite{GitHubze4:online}, on different devices. 
The experimental settings and target platforms are shown in Table \ref{tb:exp-rtos-set} and Table \ref{tb:device-info}. 
In the experiment, multiple sample applications from the Zephyr project are collected for a comprehensive evaluation. 
For each sample firmware, we instrument all of its instructions with a batch size of 1000. We use IDA \cite{IDApro:online} to extract the address of each instruction from the compiled firmware and instrument them using PIFER with \cmne{NOP} payload. An external debugger is used to reflash the patched firmware into the target device and monitor the execution. When the program fails to execute correctly it will fall back to the original Hard Fault handler, which can be confirmed by a debugger script.

\begin{table}[t]
    \centering
    \small
    \begin{threeparttable}
    \caption{The evaluation settings.}
    \label{tb:exp-rtos-set}
    \setlength{\tabcolsep}{5mm}{
    \begin{tabular}{cc}
        \toprule
        Item &  Setting       \\ 
        \midrule
        OS Version  & Zephyr 3.2.99    \\ 
        Toolchain  & GNU Arm Embedded Toolchain 12.2.0  \\ 
        \multirow{2}{*}{Applications}   & Blinky, Button, Basic Thread Example, \\ 
        &  File system shell, Logging, TinyCrypt    \\ 
        \bottomrule
        \end{tabular}
    }
    \end{threeparttable}
\end{table}

\begin{table}[t]
    \centering
    \small
    \begin{threeparttable}
    \caption{Target platforms}
    \label{tb:device-info}
    
    \begin{tabular}{cccc}
        \toprule
        Device MCU & Architecture    & Frequency & RAM/FLASH   \\ 
        \midrule
        STM32L073  & Cortex-M0  & 32MHz      & 20KB/192KB      \\ 
        STM32F103  & Cortex-M3  & 72MHz      & 20KB/64KB     \\ 
        nRF52840   & Cortex-M4  & 64MHz      & 256KB/1MB      \\ 
        LPC1549    & Cortex-M3  & 72MHz      & 36KB/256KB     \\ 
        LPC55S69    & Cortex-M33  & 150MHz     & 320KB/640KB   \\ 
        \bottomrule
        \end{tabular}
    \end{threeparttable}
\end{table}

\begin{table}[t]
  \centering
  \small
  \begin{threeparttable}
  \caption{Instructions that PIFER cannot instrument.}
  \label{tb:exp-inst-cannot}
  \setlength{\tabcolsep}{5mm}{
  \begin{tabular}{cc}
      \toprule
      Instruction &  Description       \\ 
      \midrule
      BKPT  & Breakpoint    \\ 
      CPSIE/CPSID  & Disable Interrupts/Enable Interrupts  \\ 
      MRS/MSR   & Read from/Write to special register    \\ 
      SEV    & Send Event    \\ 
      SVC    & Supervisor Call    \\ 
      WFE    & Wait For Event    \\ 
      WFI    & Wait For Interrupt    \\ 
      \bottomrule
      \end{tabular}
  }
  \end{threeparttable}
\end{table}

\noindent \textbf{Result and analysis.} The results show that the overall correct instrumenting rate of PIFER was 98.9\% on all platforms. Specifically, the instructions that PIFER could not handle are listed in Table \ref{tb:exp-inst-cannot}. 
In general, these instructions also work through interrupts/exceptions but cannot preempt the priority of the hook system. 
Nevertheless, as all these instructions have dedicated functions and are not typically the target of instrumenting for security purposes, their impact on the practicality of PIFER is negligible.

\subsection{Measurement of the Overhead (Correctness, Performace)}

\noindent \textbf{Experiment settings.} CoreMark \cite{coremark:online} is a standardized benchmark suite designed to measure the performance of microcontrollers using a set of representative workloads, including matrix operation, state machine, linked list manipulation and CRC checksum calculation. Our target platform is the LPC55S69 from Table \ref{tb:device-info}. The CoreMark implementation followed the official guide \cite{CoremarLPC:online} and is also open-sourced in our GitHub repository.  
The device under test will pull up the output levels of two pins before and after the benchmark test, with an oscilloscope to measure the time. In each experiment, we use PIFER to randomly instrument $N$ instructions in the firmware with an empty payload, and a counter is integrated into the hook system to monitor the number of received traps.

\begin{figure}[t]
    \centering
    \begin{subfigure}[b]{0.80\linewidth}
        \centering
        \includegraphics[width=1.0\linewidth]{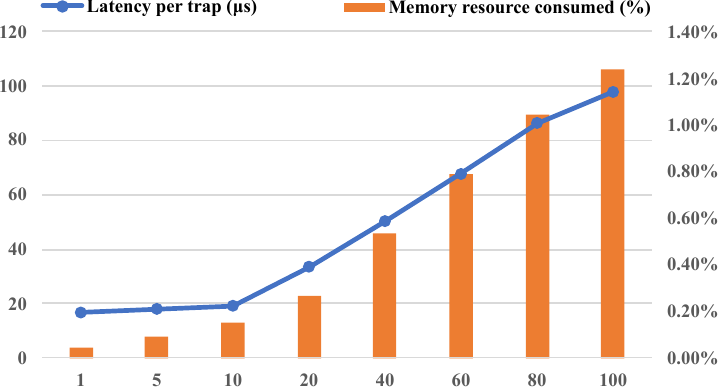}
        \caption{Small scale: $N$=1-100}
        \label{fig:coremark-result-100}
    \end{subfigure}
    \\
    \begin{subfigure}[b]{0.80\linewidth}
        \centering
        \includegraphics[width=1.0\linewidth]{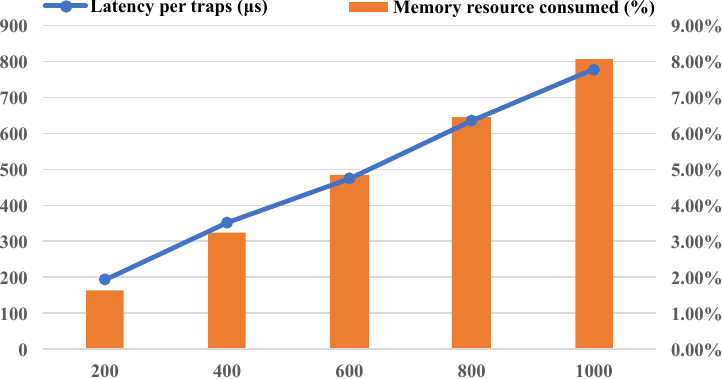}
        \caption{Larger scale: $N$=200-1000}
        \label{fig:coremark-result-1000}
    \end{subfigure}
       \caption{CoreMark experiment results.}
       \label{fig:nRF52840}
\end{figure}

\noindent \textbf{Result and analysis.} The results are shown in Figure \ref{fig:coremark-result-100}. The blue line represents the average latency caused by processing each trap, obtained by dividing the total time overhead by the number of received traps. 
The bar chart represents extra memory consumed by the instrumentation, including the increase of the binary firmware size and the stack space occurred by the hook system. 
From the results, we find that the performance loss of the PIFER rises with the number of instrumented instructions, and each additional hooking point will bring an overhead of $\sim 1\mu$s. We attribute this phenomenon to the $\mathcal{O}(N)$ lookup complexity in our dispatcher. Implementing a more efficient algorithm could help improve this issue, but requires trade-offs between time and space overheads. In general, PIFER has the ability to correctly instrument the firmware at thousands of addresses. For tasks requiring only dozens of instrumenting points at certain functions, such as vulnerability patching or performance profiling, the overhead of PIFER is negligible.

\subsection{Case Study: Real-world Firmware Customization (Correctness, Scalability)}

\begin{figure}[t]
  \centering
  \begin{subfigure}[b]{0.33\linewidth}
      \centering
      \includegraphics[width=1.0\textwidth]{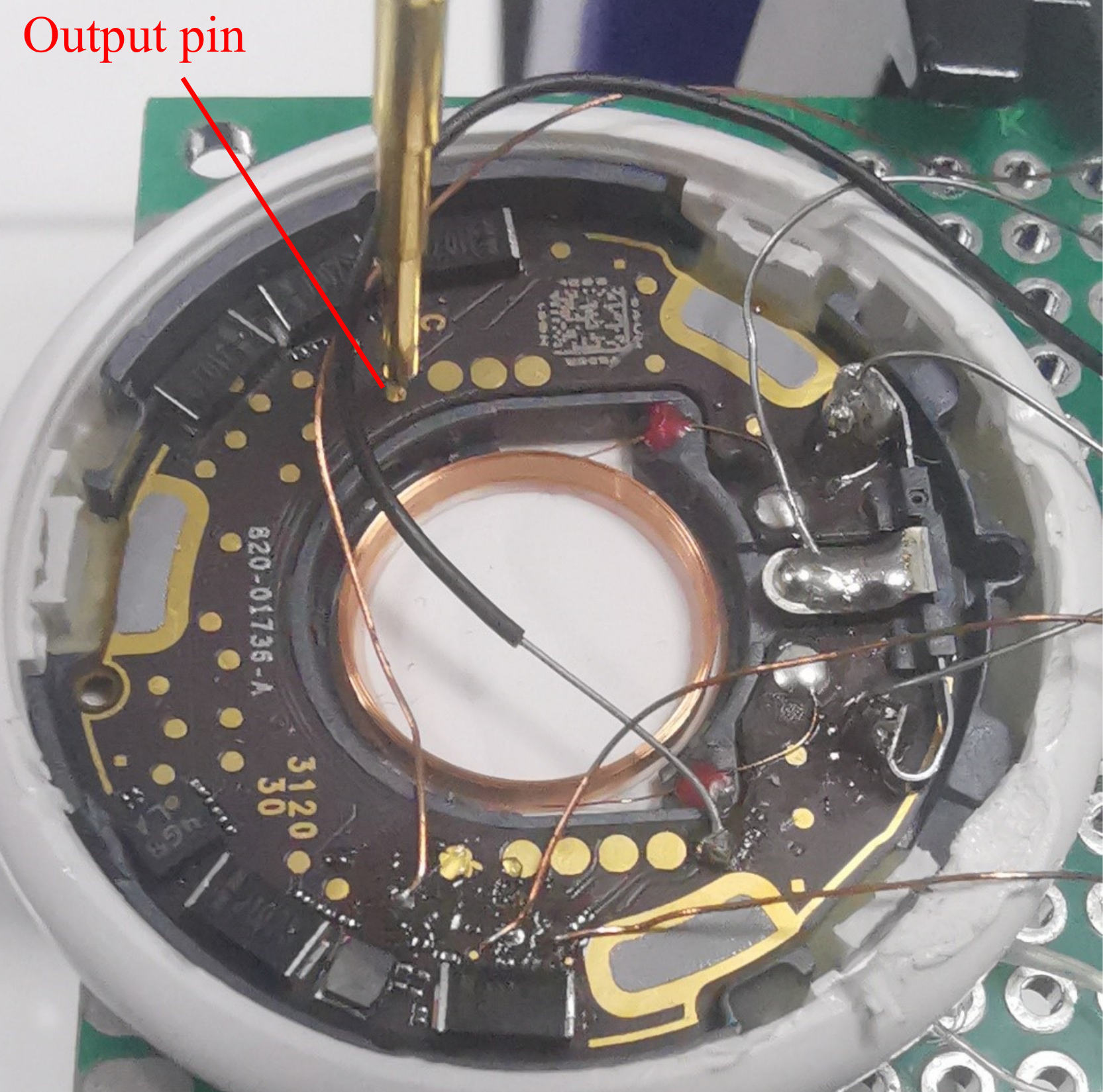}
      \vspace{0.01pt}
      \caption{The pin adopted.}
      \label{fig:airtag-pin}
  \end{subfigure}
  \begin{subfigure}[b]{0.66\linewidth}
      \centering
      \includegraphics[width=1.0\textwidth]{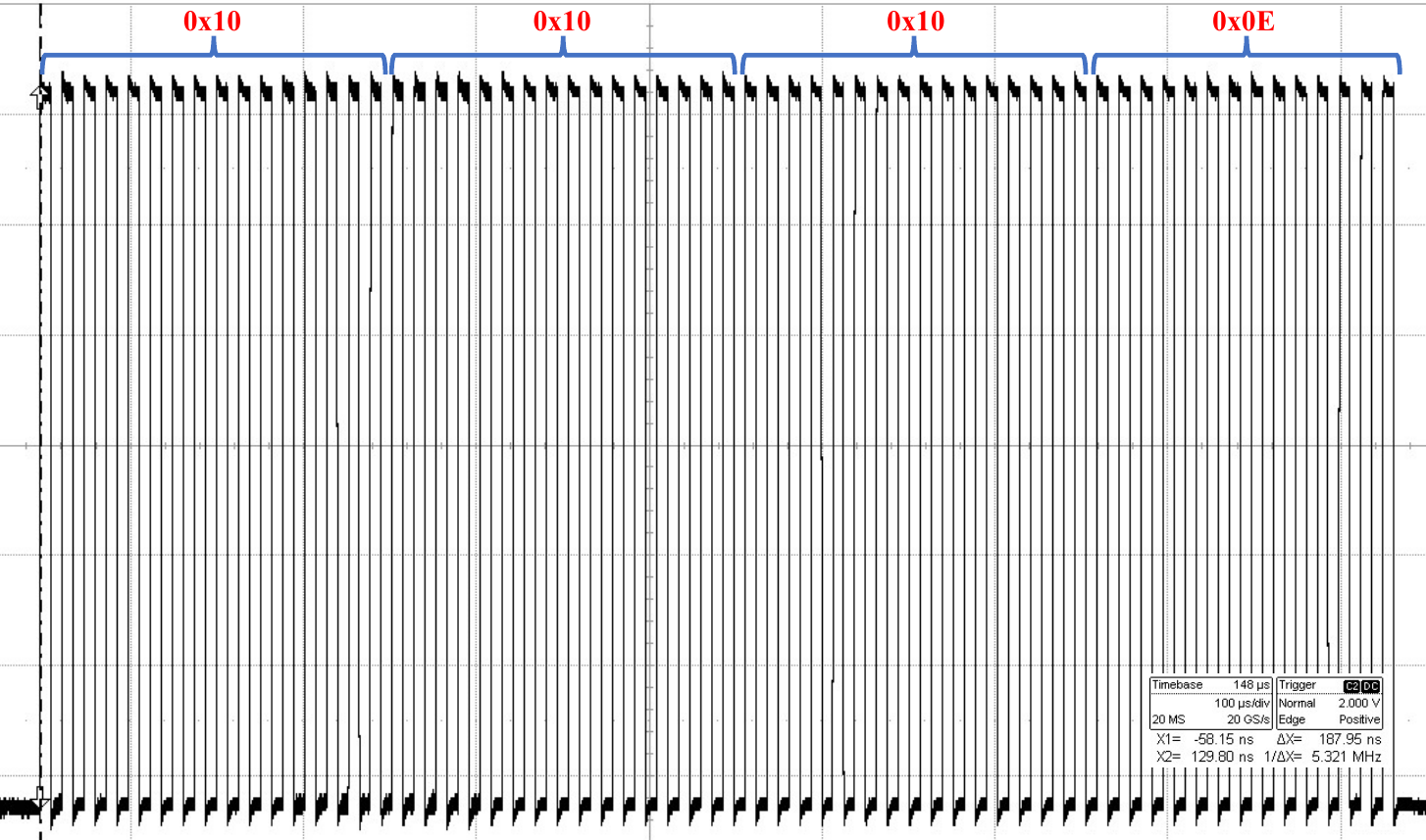}
      \caption{Leaking the first byte of LTK (0x3E).}
      \label{fig:airtag-output}
  \end{subfigure}
     \caption{Experiment with Airtag.}
     \label{fig:airtag}
\end{figure}

\noindent \textbf{Firmware acquisition and basic analysis.} To further demonstrate PIFER's ability to handle real-world complex firmware, we apply it to the Apple Airtag \cite{AirTagAp82:online}, a product for which no source code is available. For the sake of reproducibility, our experiments were conducted with a brand-new device with firmware version 1.0.225. We gain the dump and flash ability on Airtag using a public glitch tool \cite{GitHubst27:online} and extract the binary firmware\footnote[1]{Note that we are not trying to "attack" the Airtag, but verifying PIFER's ability to handle unknown firmware. Thus, the glitch tool itself is not our focus.}. 
After a manual analysis, we identify the location of the EVT and recover some of the application logic. 
To verify PIFER's ability for firmware customization, we apply it to insert a proof-of-concept protocol into the firmware, which leaks the long-term key (LTK) for Bluetooth communication\cite{scarfone2008guide} through a pin shown in Fig. \ref{fig:airtag-pin}.

\noindent \textbf{Results.} After patching the key leak function into the firmware and re-flashing it back to an Airtag, we connect the corresponding pin to an oscilloscope to visualize the output. As shown in Fig. \ref{fig:airtag-output}, the first byte of LTK was successfully encoded in the voltage level, proving the practicality and correctness of PIFER.

\section{Limitations and Future Work}

\noindent \textit{More robust instruction translation.} As noted in Section \ref{sec:proxy-rewriter}, although no samples are observed in real-word firmware, the current free register searching has corner cases in theory. Improvements to this scheme are left to future work, such as identifying usable registers with advanced program analysis.

\noindent \textit{Improvements in dispatchers.} We currently use an $\mathcal{O}(n)$ lookup for finding the handler for a specific address, and the performance will degrade as the number of hooks increases. In the future, we can explore more efficient implementations.

\section{Conclusion}

In this paper, we present the design of PIFER, a static binary instrumenting tool for bare-metal embedded firmware. 
By abusing the exception handling mechanism, PIFER builds a self-contained hooking system providing a fine-grained instrumenting functionality. 
We also propose an instruction translating scheme to overcome the unrecoverable modifications in FLASH code, ensuring that the firmware works correctly after patching. 
Moreover, we implement and open source a full prototype of PIFER on the ARM architecture. 
Comprehensive experiments confirm PIFER's capacity to process complex, real-world targets with acceptable overhead, demonstrating the effectiveness and practicality of our work.

\bibliographystyle{ACM-Reference-Format}
\bibliography{base}

\end{document}